\documentclass[superscriptaddress,reprint,amsmath,amssymb,aps,prx,showpacs,floatfix]{revtex4-1}

\usepackage[dvips]{color,graphicx}
\usepackage[usenames,dvipsnames,svgnames,table]{xcolor}
\DeclareGraphicsExtensions{.eps, .jpg, .png, .bmp}

\newcommand*{\ang}{\mbox{\normalfont\AA}}
\newcommand*{\eV}{\, \textrm{eV}}
\newcommand*{\meV}{\, \textrm{meV}}
\newcommand*{\eVmolecule}{\, \textrm{eV}/\textrm{H}_2\textrm{O}}

\begin{document}

\title{Vibrational renormalisation of the electronic band gap in hexagonal and cubic ice}

\author{Edgar~A.~Engel}
\email{eae32@cam.ac.uk}
\affiliation{TCM Group, Cavendish Laboratory, University of Cambridge,
J. J. Thomson Avenue, Cambridge CB3 0HE, United Kingdom}
\author{Bartomeu~Monserrat}
\affiliation{TCM Group, Cavendish Laboratory, University of Cambridge,
J. J. Thomson Avenue, Cambridge CB3 0HE, United Kingdom}
\affiliation{Department of Physics and Astronomy, Rutgers University, 
Piscataway, New Jersey 08854-8019, USA}
\author{Richard~J.~Needs}
\affiliation{TCM Group, Cavendish Laboratory, University of Cambridge,
J. J. Thomson Avenue, Cambridge CB3 0HE, United Kingdom}

\date{\today}

\begin{abstract}
  Electron-phonon coupling in hexagonal and cubic water ice is studied using 
  first-principles quantum mechanical methods.  We consider 29 distinct 
  hexagonal and cubic ice proton-orderings with up to 192 molecules in the 
  simulation cell to account for proton-disorder. We find quantum zero-point 
  vibrational corrections to the minimum electronic band gaps ranging from 
  $-1.5$ to $-1.7 \eV$, which leads to improved agreement between calculated 
  and experimental band gaps. Anharmonic nuclear vibrations play a negligible 
  role in determining the gaps. Deuterated ice has a smaller band-gap 
  correction at zero-temperature of $-1.2$ to $-1.4 \eV$. Vibrations reduce 
  the differences between the electronic band gaps of different proton-orderings 
  from around $0.17 \eV$ to less than $0.05 \eV$, so that the electronic band 
  gaps of hexagonal and cubic ice are almost independent of the proton-ordering 
  when quantum nuclear vibrations are taken into account.
  The comparatively small reduction in the band gap over the temperature 
  range 0--240 K of around $0.1 \eV$ does not depend on the proton ordering, 
  or whether the ice is protiated or deuterated, or hexagonal or cubic.  
  We explain this in terms of the atomistic origin of the strong 
  electron-phonon coupling in ice.
\end{abstract}

\maketitle

\section{Introduction}
\label{Introduction}

Ice is a key constituent of the Earth's crust and mantle
\cite{weingaertner,liebscher}, while on its surface and in its
atmosphere it plays a crucial role in the water cycle and in determining
climate \cite{bartels,baker,young}.  Accordingly, ice is one of the
experimentally and computationally most extensively studied condensed 
matter systems and its properties have been investigated across a wide 
range of temperatures and pressures \cite{bartels}.

The electronic bandstructure of ice plays an important role, e.g., in the 
redox reactions of atmospheric chemistry \cite{george,betterton,
  grannas2,heger,takenaka1,takenaka3,park,boxe,kahan,klanova},
in glaciers \cite{bishop}, Earth's interior \cite{liebscher,chandler,marcus}, 
and in structural determinations using electrochemical scanning tunnelling 
microscopy in aqueous environments \cite{cucinotta}.  The insulator to 
metal transition induced by band gap closure \cite{hama,cavazzoni,mattsson,
hermann_high_pressure_ices,pickard,galli} in high pressure ice phases is 
thought to be important in astrophysical contexts such as in the sources 
of the magnetic fields of Uranus and Neptune \cite{ness}.


\begin{figure}
   \centering
	\includegraphics[width=0.5\textwidth]{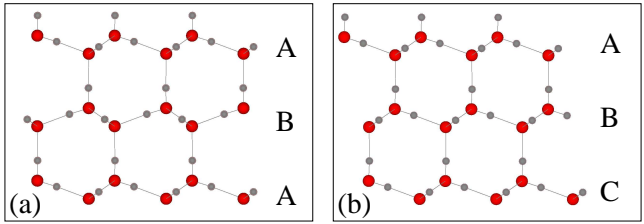}
	\caption{The oxygen sublattices of (a) hexagonal, Ih, and (b) cubic ice, 
		Ic, are shown in red. The puckered layers have ABAB and ABC stackings 
		in Ih and Ic, respectively. 
		Arbitrarily chosen particular proton-orderings obeying the Bernal-Fowler 
	ice rules are shown in grey.}
	\label{fig:Stacking}
\end{figure}
At the pressures found on Earth's surface only stable hexagonal (Ih) and 
meta-stable cubic ice (Ic) \cite{engel} (see Fig.~\ref{fig:Stacking}) 
occur naturally. The latter typically contains many stacking faults.
While experimental estimates of the zero-temperature electronic 
quasi-particle band gap, $E_{g} (0)$, in Ih of $8.8 \pm 0.4 \eV$ 
exist (see Refs.\ \cite{hahn,fang,warren,minton,painter,seki,shibaguchi}
and references therein), reliable values for Ic do not, since 
high purity samples of Ic have not been synthesised.  

First-principles electronic structure calculations can be used to study 
electronic band structures, and quasiparticle and excitonic band gaps. 
However, density functional theory (DFT) using semi-local functionals 
-- the workhorse of modern day electronic structure calculations -- normally 
underestimates band gaps. 
Much progress has been made in calculating electronic band gaps using more
accurate but computationally demanding hybrid functionals \cite{hse06}, 
screened exchange functionals \cite{screened_dft_clark}, perturbation 
theories such as the $GW$ approximation \cite{hedin_GW,hedin_lundqvist_GW,
hybertsen,godby2,aulbur_GW}, and quantum chemical \cite{cramer_QC,shavitt_QC} 
and quantum Monte Carlo techniques \cite{ceperley_alder_QMC,
foulkes_QMC,needs_QMC,drummond,williamson}. 
$GW$ calculations for proton-ordered hexagonal ice (XIh), for example,
reproduce the experimentally observed optical absorption spectra of Ih
reasonably well in comparison to semi-local DFT \cite{hahn,fang}. 
However, $GW$ results for hexagonal ice of $9.6 \pm 0.4 \eV$ 
\cite{hahn,fang} overestimate $E_{g}$ by around $1 \eV$.

Nuclear vibrations and proton-ordering are often neglected in studying 
band gaps of water and ice, but these effects have remarkable consequences 
for the electronic bandstructure (\cite{galli,monserrat} and this work).  
Measurements of band gaps of water provide strong experimental evidence for 
the importance of nuclear vibrations \cite{sanchez_valle}. In liquid water, 
$E_{g}$ depends strongly on temperature which indicates large thermal 
vibrational effects and suggests large zero-point (ZP) quantum vibrational 
effects \cite{sanchez_valle,galli,monserrat_exp}.  This observation motivates 
us to investigate whether vibrational effects could play a significant role 
in explaining the discrepancy between the values of $E_{g}$ obtained within 
the $GW$ approximation and experiment.  We have previously reported strong 
electron-phonon coupling in various molecular crystals, amongst them a 
proton-ordering of Ih with $P6_3cm$ symmetry \cite{monserrat}.
Benchmarking computational models against experimentally accessible systems, 
such as ice under ambient pressure, is essential. Ice displays a range of 
important and interesting phenomena related to configurational disorder and 
strong effects from nuclear vibrations due to the light hydrogen nuclei, 
which must both be included to obtain accurate atomistic simulations.

In this work we focus on vibrational corrections to the electronic band gaps 
of Ih and Ic, accounting for quantum nuclear vibrations and thermal effects 
(Sec.\ \ref{BGapCorr}). We investigate the effects of replacing the hydrogen 
atoms in Ih and Ic by heavier deuterons (Sec.\ \ref{HeavyIce}), evaluate the 
role of vibrational anharmonicity (Sec.\ \ref{Anharmonicity}), and determine 
the atomistic origins of the strong electron-phonon coupling (Sec.\ \ref{Mechanistics}).

\section{Computational Model}
\label{Methods}
\subsection{Stacking faults and proton-disorder}
\label{ProtonDisorder}
It was long believed that ice at ambient pressure occurs in the
hexagonal (Ih) and cubic (Ic) forms shown in Fig.~\ref{fig:Stacking}.
Both Ih and Ic consist of tetrahedrally coordinated water molecules
satisfying the ``Bernal-Fowler ice rules'' \cite{bernal_fowler}. Ih
and Ic have very similar free energies \cite{engel} and their structures differ
only in the stacking of the atomic layers (see Fig.\ \ref{fig:Stacking}).
Thermodynamically stable Ih occurs naturally in abundance and plays a
key role in determining Earth's climate \cite{bartels,baker,young}.
The metastable, ideal Ic form \cite{konig} was also thought to play 
a role in nature, but to be very rare \cite{murray_formation_ic,
shilling_vapour_pressure_ic_clouds}.
However, real ``cubic ice'' typically contains many stacking faults 
and is therefore now referred to as stacking-disordered ice (Isd).
Isd refers to the infinite set of possible stacking sequences, 
which smoothly connects Ih as one end member to Ic as the other.
Isd is a highly complex material \cite{malkin,kuhs_stacking_disorder,carr}
and is thought to play a key role in ice nucleation, a process central
to climate.
A full understanding of Isd will require knowledge of the properties 
of Ih and Ic, which we limit ourselves to in this study.

Ih and Ic are proton-disordered systems with an extensive ground-state
degeneracy that leads to Pauling's residual configurational entropy
\cite{pauling,tajima,jackson}.  Atomistic simulations of Ih and Ic
treat the hydrogen atoms explicitly and are based on representative
sets of energetically quasi-degenerate, proton-ordered structures.
The number of such structures allowed by the ice rules increases 
exponentially with the size of the simulation cell.
For large systems, the configurational entropies of Ih and Ic are 
found to be almost identical \cite{nagle_lattice_statistics,
ramirez_configurational_entropy}, and therefore they do not affect 
the relative stability.

We consider 16 distinct proton-ordered eight-molecule Ih configurations 
as constructed by Hirsch and Ojam\"ae \cite{hirsch}, and $11$ distinct 
proton-ordered eight-molecule Ic configurations \cite{raza}. We also 
consider the ``conventional'' hexagonal, 12-molecule $P6_{3}cm$ Ih and 
quasi-cubic, eight-molecule $P4_{3}$ Ic structures (numbers $13$ and 
$1$ in Figs.\ \ref{fig:Polytypes_BandGaps} (a) and (b), respectively).
We calculate differences in static lattice energies for these sets of 
Ih and Ic proton-orderings which are in good agreement with those in 
Refs.~\cite{hirsch,lekner_hydrogen_ordering_energetics}.
More importantly, the variations in static lattice energies across the 
Ih proton-orderings agree with those calculated using first principles 
methods in Ref.~\cite{singer} for a much larger set of 16 8-molecule 
orthorhombic, 14 12-molecule hexagonal and 63 48-molecule orthorhombic 
Ih proton-orderings.
These results strongly suggest that our sets of proton-orderings 
provide a good representation of proton-disordered Ih and Ic.

\subsection{Static lattice calculations}
\label{StaticLatticeCalculations}
While DFT with semi-local exchange-correlation functionals underestimates 
static electronic band gaps, it performs well in evaluating phonon 
spectra and vibrational band-gap corrections in ice, see 
Supplementary Section V \cite{supplementary} for details. We performed 
electronic structure calculations using plane-wave pseudopotential DFT 
as implemented in the {\sc castep} code \cite{clarkSPHPRP05} (version 
7.02).  We employed the Perdew-Burke-Ernzerhof (PBE) \cite{perdew_1996_PBE, 
santra_xcfunctionals_ice} semilocal generalised gradient approximation 
functional, and on-the-fly generated ultrasoft pseudopotentials 
\cite{Vanderbilt90} with core radii of $0.7\ \ang$ and $0.8\ \ang$ for 
the hydrogen and oxygen atoms, respectively.  We used a plane-wave 
energy cut-off of $1600 \eV$ and Monkhorst-Pack electronic reciprocal space 
grids of spacing less than $2\pi \times 0.04\ \ang^{-1}$ for the total energy 
calculations and geometry optimisations 
involved in the vibrational calculations described in Section~\ref{VibrationalCalculations}.  
The resulting energy differences between frozen-phonon configurations are 
converged to within $10^{-4}\eVmolecule$, the atomic positions are converged 
to within $10^{-5}\ang$, and the residual forces to within $10^{-4}\eV/\ang$.
We used a lower plane-wave energy cut-off of $800 \eV$ for the band structure 
calculations of frozen-phonon structures, which is sufficient for accurate 
electron-phonon coupling calculations as described in Section~\ref{ElPhCoupling}.

\subsection{Vibrational calculations}
\label{VibrationalCalculations}
We obtained the harmonic vibrational normal modes, which define the normal 
phonon coordinates $q_{n{\bf k}}$ and the harmonic vibrational frequencies
$\omega_{n{\bf k}}$ of a vibrational mode $(n, {\bf k})$, using a finite 
displacement method \cite{kune}.
$n$ and ${\bf k}$ denote the branch index and vibrational Brillouin Zone 
wavevector of the vibrational mode, respectively. The phonon coordinates 
$q_{n{\bf k}}$ denote the atomic displacements along the vibrational mode 
$(n, {\bf k})$.
Anharmonic vibrations were calculated using the method described in
Ref.\ \cite{anharmonic_method}. We investigate the 3$N$-dimensional BO
energy surface (where $N$ is the number of atoms in the simulation
cell) by mapping 1D subspaces along the harmonic normal mode axes. 
Using 11 equally spaced sampling points for each 1D subspaces was found 
to lead to converged results.
Large maximum amplitudes of four times the harmonic root-mean-square 
(RMS) displacements were chosen to accurately describe the BO energy 
surface where anharmonicity is important. 
We then construct the 3$N$-dimensional BO surface from the 1D subspaces 
which were fitted using cubic splines \cite{engel}.
The anharmonic vibrational Schr\"{o}dinger equation was solved within
a vibrational self-consistent field (VSCF) framework. 
The anharmonic 
vibrational wave function was expanded as a Hartree product, 
\begin{equation}
\left\vert \phi_{{\bf S}} ({\bf q}) \right\rangle = 
\prod_{n,{\bf k}} \left\vert \varphi_{S_{n{\bf k}}} (q_{n{\bf k}}) \right\rangle \, ,
\end{equation}
of single-particle anharmonic eigenstates, $\left\vert \varphi_{S_{n{\bf k}}} 
(q_{n{\bf k}}) \right\rangle$, with energy ${\textrm E}_{S_{n{\bf k}}}$. 
${\bf S}$ denotes the vibrational eigenstate whose elements $S_{n{\bf k}}$ 
label the states of the vibrational modes $(n, {\bf k})$. In practice, 
we expand the anharmonic states $\left\vert \varphi_{S_{n{\bf k}}} (q_{n{\bf k}}) 
\right\rangle$ in a basis of simple harmonic oscillator eigenstates.
The inclusion of 25 simple harmonic oscillator states for each 
degree of freedom was found to lead to converged results (see Supplementary 
Section I).

\subsection{Electron-phonon coupling}
\label{ElPhCoupling}
Within the Born-Oppenheimer approximation, the vibrationally corrected
band gap, $E_{g}(T)$, at temperature $T$ may be written as 
\begin{equation}
E_{g}(T) = \frac{1}{Z(T)} \sum_{{\bf S}} \left\langle \phi_{{\bf S}} ({\bf q}) \right\vert 
{\hat E}_{g}({\bf q}) \left\vert \phi_{{\bf S}} ({\bf q}) \right\rangle 
\exp{ \left(-\beta {\textrm E}_{{\bf S}}\right) },
\end{equation}
where 
\begin{equation}
Z(T) = \sum_{{\bf S}} \left\langle \phi_{{\bf S}} ({\bf q}) \vert 
\phi_{{\bf S}} ({\bf q}) \right\rangle \exp{ \left(-\beta {\textrm E}_{{\bf S}}\right) } 
\, , \,
\beta = 1/k_{\textrm B} T
\end{equation}
is the partition function, ${\hat E}_{g}({\bf q})$ is the gap for a
frozen phonon structure with atomic positions ${\bf q} = (\ldots, 
q_{n{\bf k}}, \ldots)$, and $\left\vert \phi_{{\bf S}} ({\bf q})
\right\rangle$ is the harmonic or (where explicitly stated) anharmonic 
vibrational eigenstate with energy ${\textrm E}_{{\bf S}}$. 
The summation over vibrational eigenstates ${\bf S}$ includes the 
vibrational ground-state and vibrationally excited states. 
The above theory provides a semiclassical approximation to the change 
in the band gap arising from nuclear motion \cite{patrick}.

In this work we use a quadratic approximation to ${\hat E}_{g}({\bf q})$, 
which allows us to asses the importance of individual vibrational modes,
to investigate microscopic mechanisms, and a Monte Carlo sampling method 
to calculate accurate numerical values of $E_{g}(T)$.  In the quadratic 
approximation ${\hat E}_{g}({\bf q})$ is approximated as
\begin{equation} 
	{\hat E}_{g}({\bf q}) = {\hat E}_{g}({\bf 0}) + \sum_{n,{\bf k}} 
	a_{n{\bf k}} q_{n{\bf k}}^2 \, .
\end{equation}
The harmonic vibrational wave function is
symmetric and odd terms in the polynomial expansion of ${\hat E}_{g}({\bf q})$
vanish when evaluating the expectation value. Hence, the quadratic term is 
the first non-zero correction to the static band gap and errors are of 
${\cal O}({\bf q}^4)$. We calculate the diagonal quadratic coefficients, 
$a_{n{\bf k}}$, using a frozen phonon method, displacing the atoms by 
$\Delta q_{n{\bf k}}$ along phonon modes, and setting 
$a_{n{\bf k}} = ({\hat E}_{g} (\Delta q_{n{\bf k}}) + 
{\hat E}_{g}(-\Delta q_{n{\bf k}})) / (2 \Delta q_{n{\bf k}}^2)$.  
We note that Eq.\ (4) is equivalent to the Allen-Heine-Cardona theory
including off-diagonal Debye-Waller terms \cite{gonze}. 

\begin{figure}
	\centering
	\includegraphics[width=0.4\textwidth]{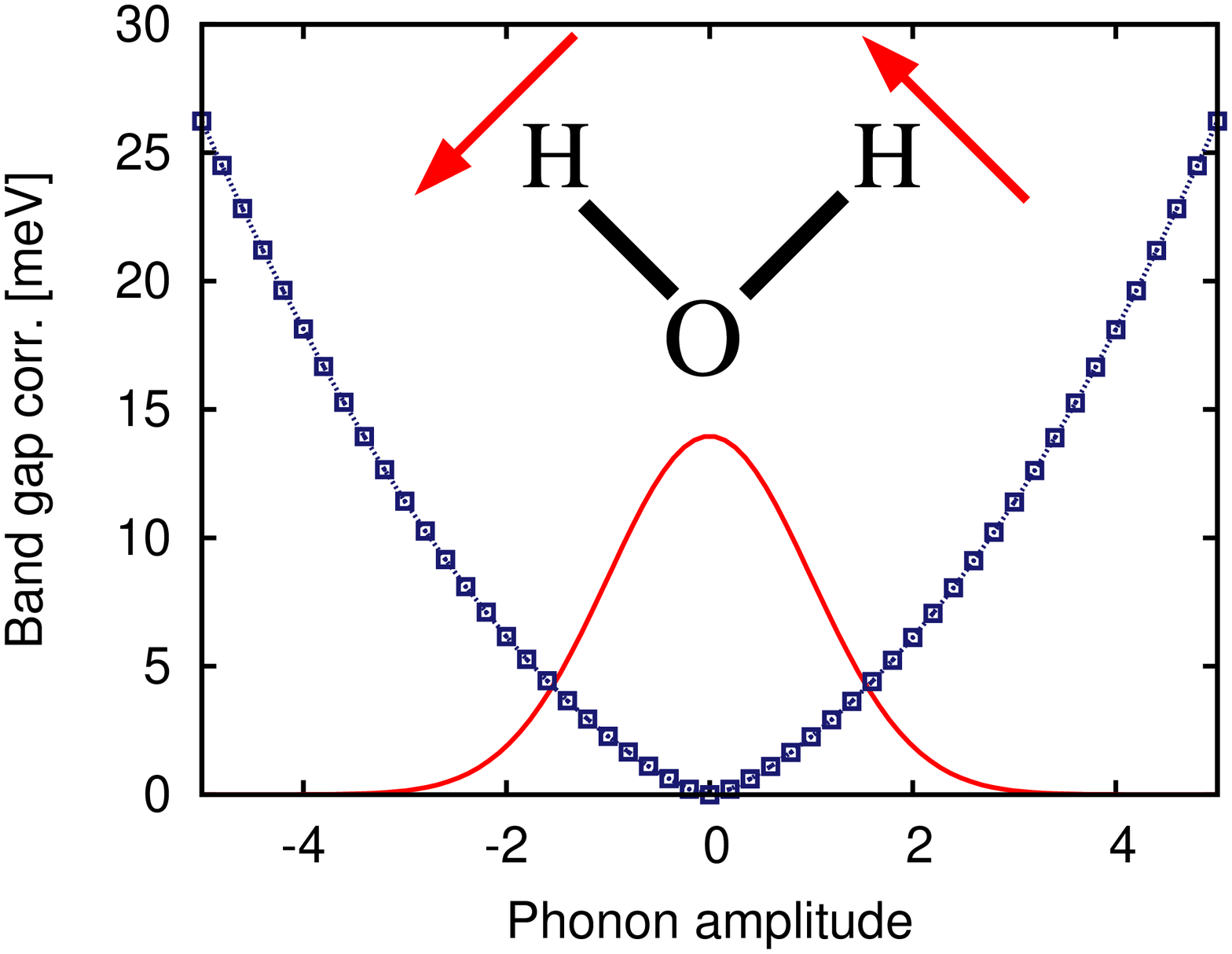}
	\includegraphics[width=0.4\textwidth]{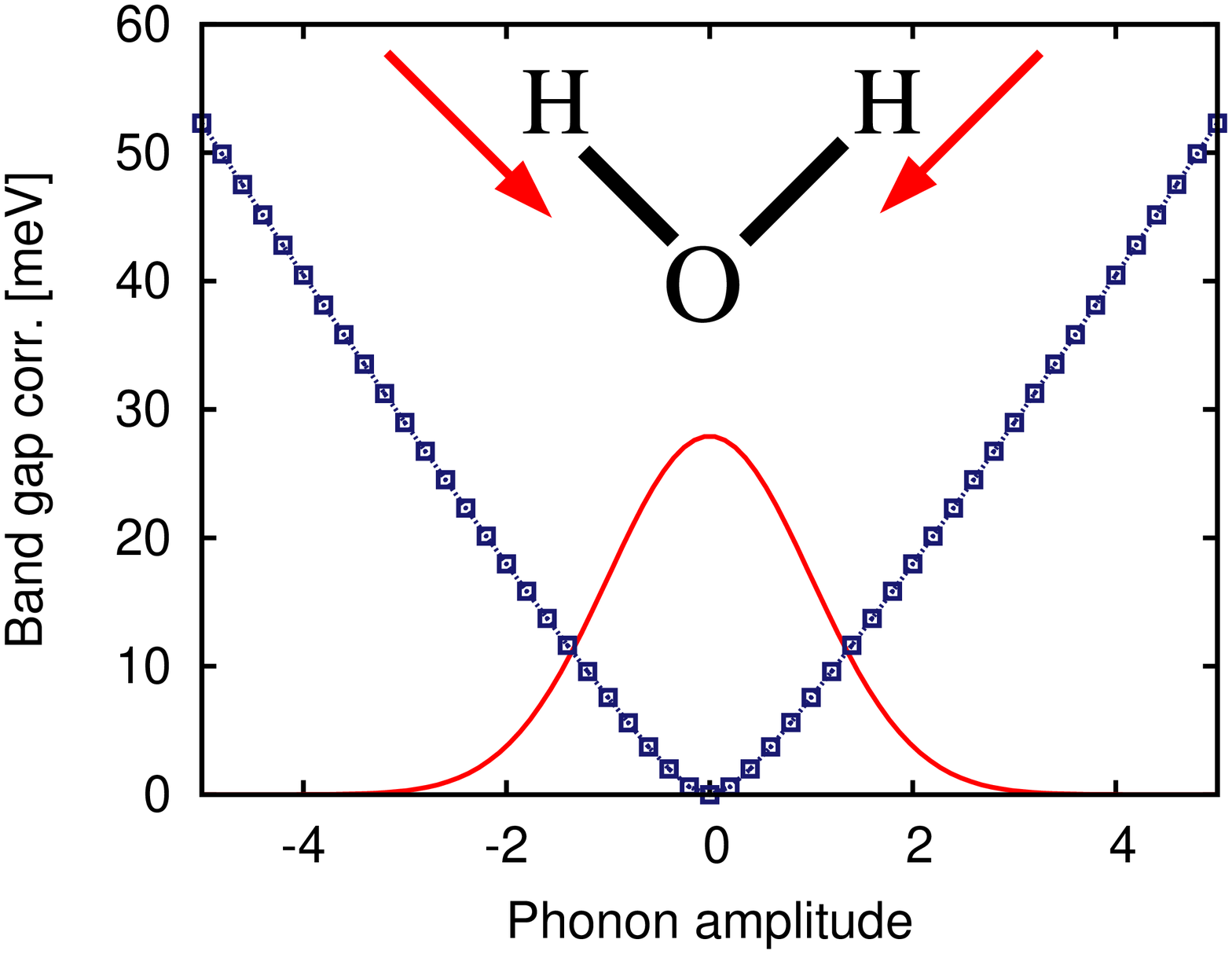}
	\caption{Failure of the quadratic approximation for electron-phonon 
		coupling for a librational (top) and a high frequency molecular 
		symmetric O-H stretching mode (bottom). 
		The strongly non-quadratic dependence of $\Delta {\hat E}_{g} ({\bf q})$ 
		on the atomic positions ${\bf q}$ (blue) is particularly common for the 
		librational and molecular modes of water ice. The harmonic vibrational 
	density is shown in red.}
	\label{fig:FailureHarmApprox}
\end{figure}

Although the quadratic approximation is valuable in obtaining atomistic 
insights, it does not produce accurate values of $E_{g}(T)$ for Ih and Ic. 
This arises because ${\hat E}_{g}({\bf q})$ is generally not well described 
by a quadratic form, which also implies that the widely used Allen-Heine-Cardona 
theory is insufficient for studying electron-phonon coupling in ice.
Examples of vibrational modes with particularly non-quadratic behaviour 
are shown in Fig.\ \ref{fig:FailureHarmApprox}.
Note that the non-quadratic dependence of the band gap on the
atomic displacements is distinct from anharmonicity in nuclear vibrations 
and we therefore treat them independently.

Instead of using the quadratic approximation to $\Delta {\hat E}_{g} ({\bf q})$, 
we therefore evaluate accurate values of $E_{g}(T)$ using Monte Carlo sampling, 
with $N$ frozen phonon structures, $\left\{ {\bf q}^i \right\}$, 
randomly drawn from the vibrational density, as $1/Z(T) \sum_{{\bf S}} \left\vert \phi_{{\bf S}} 
({\bf q}) \right\vert^2 \exp\left(-\beta {\textrm E}_{{\bf S}} \right)$, which gives
\begin{equation}
	E_{g}(T) = \frac{1}{N} \sum_{i=1}^N {\hat E}_{g}({\bf q}^i) .
\end{equation}
$N$ is typically larger than $500$.
The Monte Carlo approach includes all higher-order terms in ${\hat E}_{g}({\bf q})$
neglected in the quadratic approximation and, unlike the quadratic 
approximation, remains valid for an anharmonic nuclear wavefunction.

In order to obtain finite temperature band-gap corrections 
we resample band-gap corrections using the finite temperature nuclear density 
with its wider tails.

To obtain the band gap correction due to anharmonic nuclear vibrations, on 
the other hand, we employ a reweighting approach as in Ref.\ \cite{monserrat2}, 
which is efficient and accurate for ice, since anharmonicity leads to a narrower, 
more localised nuclear density distribution. We reuse the band gap samples, 
${\hat E}_{g}({\bf q}^i_{\textrm{har}})$, drawn from the harmonic vibrational 
density, $\left\vert \phi^{\textrm{har}} ({\bf q}) \right\vert^2 / Z^{\textrm{har}}(0)$, 
to calculate the band gap correction due to anharmonic nuclear vibrations as
\begin{equation}
E^{\textrm{anh}}_{g}(T) = \frac{1}{N} \sum_{i=1}^N w_i {\hat E}_{g}({\bf q}^i_{\textrm{har}}) ,
\end{equation}
where the weights, $w_i$, are calculated from the harmonic and anharmonic 
nuclear probability densities
\begin{equation}
w_i = \frac
{\sum_{{\bf S}} \left\vert \phi_{{\bf S}}^{\textrm{anh}} ({\bf q}) \right\vert^2 
\textrm{e}^{-\beta {\textrm E}_{{\bf S}}^{\textrm{anh}}} / Z^{\textrm{anh}}(T)}
{\sum_{{\bf S}} \left\vert \phi_{{\bf S}}^{\textrm{har}} ({\bf q}) \right\vert^2 
\textrm{e}^{-\beta {\textrm E}_{{\bf S}}^{\textrm{har}}} / Z^{\textrm{har}}(T)} \, .
\end{equation}
As shown in Section \ref{Anharmonicity}, including vibrational anharmonicity 
changes the vibrational band-gap correction, $\Delta E_{g}(T) = E_{g}(T) - 
{\hat E}_{g}({\bf 0})$, by only around 2 and 5 \% in Ih 
and Ic, respectively, and (even up to the melting temperature) predominantly 
via the ZP band-gap correction. 
Hence, unless stated otherwise, all results were obtained using the harmonic
approximation for the nuclear vibrations.
%

The same Monte Carlo approach could be employed to calculate vibrationally 
renormalised optical absorption spectra as demonstrated, for example, in Ref.\ 
\cite{zacharias}. For ice, excitonic effects would have to be included, for 
example, using the Bethe Salpeter equation \cite{bethe}, to obtain accurate 
optical absorption spectra.
A first approximation at a much reduced computational cost could 
be obtained by simply adding the vibrational shift, $\Delta E_{g}(T)$, 
calculated in this work to the static lattice optical absorption spectra 
of Refs.\ \cite{hahn} and \cite{fang}. 
Vibrationally renormalised optical absorption spectra of ice have not been 
calculated before, although they have been reported for other materials 
\cite{marini,kioupakis,giustino}.

\begin{figure}
	\centering
	\includegraphics[width=0.4\textwidth]{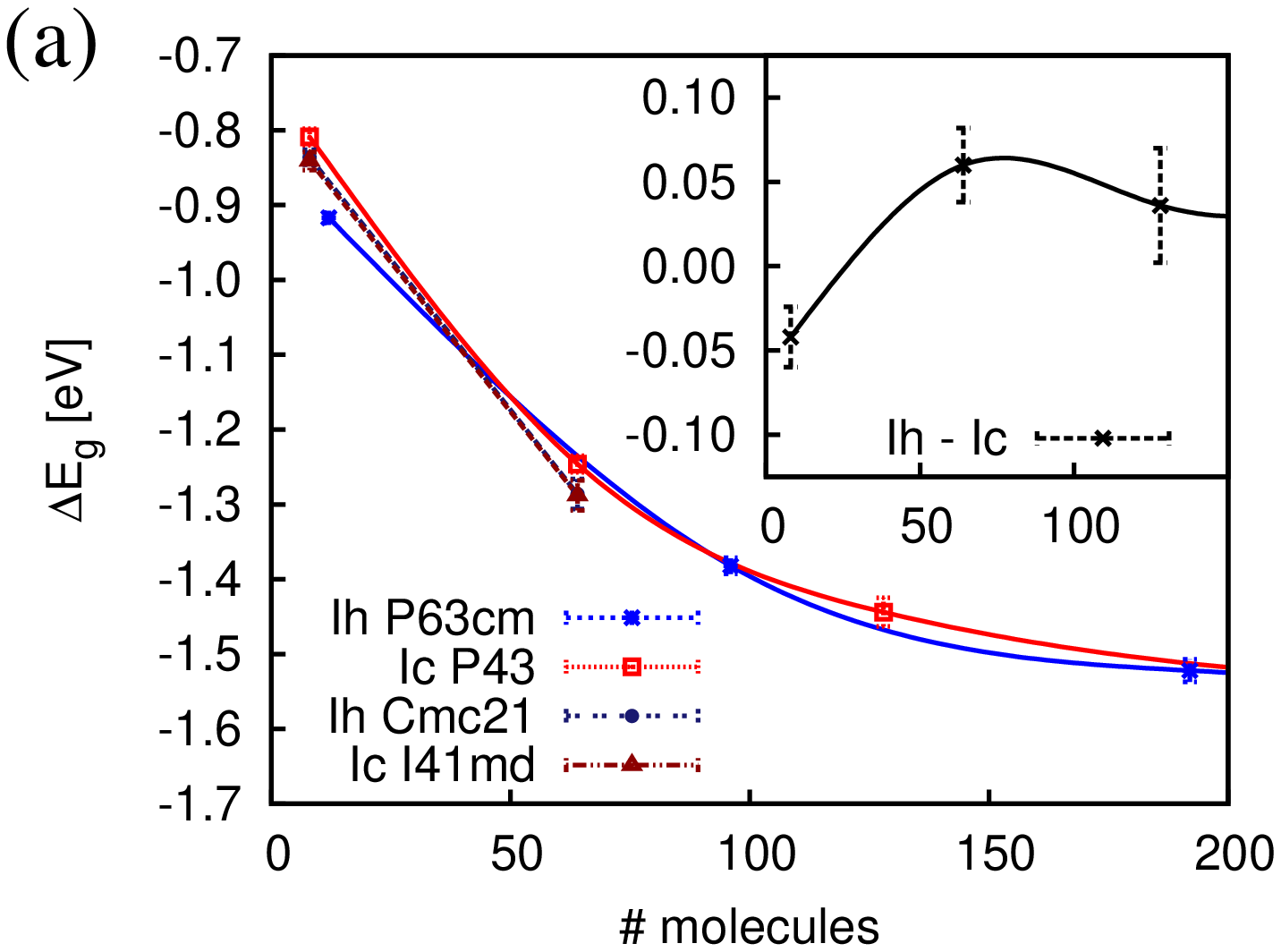}
	\includegraphics[width=0.4\textwidth]{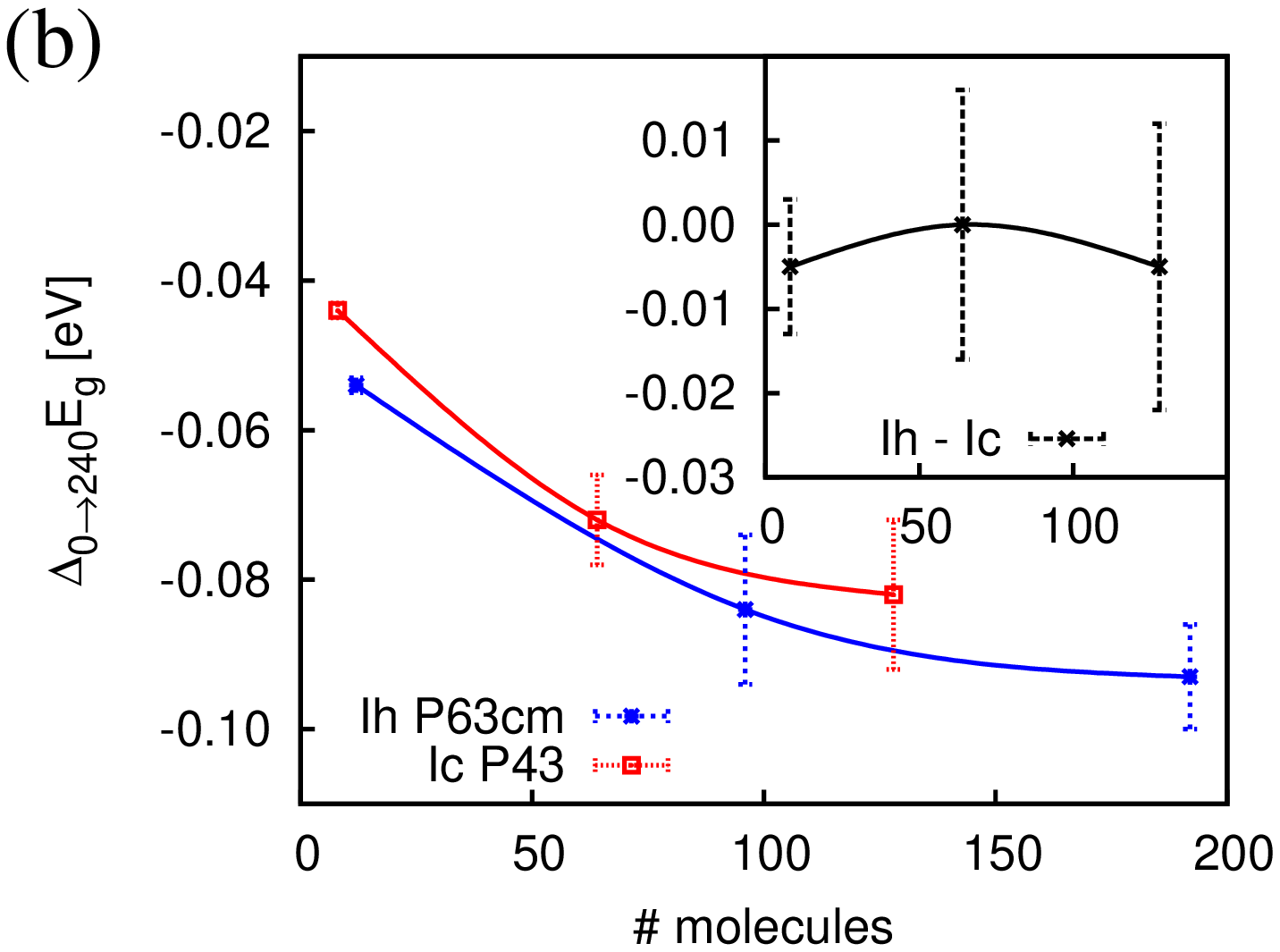}
	\caption{Convergence of (a) the ZP band gap correction, $\Delta E_{g}(0)$, 
		and (b) the thermal contribution $\Delta_{0 \mathrm{\rightarrow} T} 
		E_{g} \equiv \Delta E_{g}(T) - \Delta E_{g}(0)$ with simulation cell 
		size. 
		The difference in $\Delta E_{g}(T)$ between Ih and Ic (and 
		proton-orderings) converges rapidly with simulation cell size, even 
		though the absolute $\Delta E_{g}(T)$ do not. This can be seen in the 
		insets of (a) and (b), which show the difference in $\Delta E_{g}(0)$ 
		and in $\Delta_{0 \rightarrow 240} E_{g}(T)$ between Ih $P6_3cm$ and 
		Ic $P4_3$, respectively. Comparing $\Delta E_{g}(0)$ of Ih $P6_3cm$, 
		Ih $Cmc2_1$, Ic $P4_3$ and Ic $I4_1md$ shown in (a) provides further confirmation.
		The error bars indicate the statistical errors due to Monte Carlo sampling.}
	\label{fig:SC_Conv}
\end{figure}

The absolute sizes of the vibrational corrections, $\Delta E_{g}(T)$, 
for different proton-orderings converge very slowly with the size of 
the simulation cell or, equivalently, with the density of the Brillouin 
Zone sampling (Figs.\ \ref{fig:SC_Conv} (a) and (b), main panels). 
However, the differences in $\Delta E_{g}(T)$ between structures converge 
rapidly (Figs.\ \ref{fig:SC_Conv} (a) and (b), insets). 
The relative $\Delta E_{g}(T)$ of various orthorhombic, 8-molecule proton-orderings 
are converged to better than around $50 \meV$. Despite the difference in shape and 
size of the simulation cell, the $\Delta E_{g}(T)$ of the hexagonal, 12-molecule 
Ih $P6_3cm$ proton-ordering is converged to within $75 \meV$ relative to the 
orthorhombic, 8-molecule proton-orderings. 
This justifies the use of 8-molecule simulation cells in obtaining accurate 
relative band-gap corrections $\Delta E_{g}$ between different proton-orderings 
and between Ih and Ic. 
Supplementary Fig.\ S12 provides further support for this observation.
We use simulation cells with up to 192 molecules to converge the absolute 
$\Delta E_{g}(T)$ with respect to simulation cell size to within $50 \meV$.

\section{Vibrational Band Gap Renormalisation}
\label{BGapCorr}

\begin{figure}
	\centering
	\includegraphics[width=0.4\textwidth]{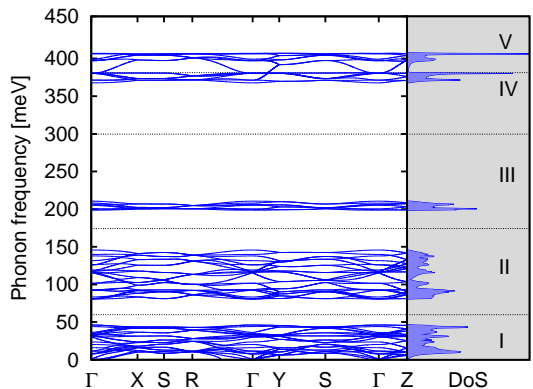}
	\caption{Vibrational dispersion (left panel) and DoS (right panel) of 
		64-molecule Ih $Cmc2_1$. The DoS splits into crystal modes (I), librational 
		modes (II), molecular bending modes (III) and molecular anti-symmetric (IV) 
		and symmetric (V) O-H stretching modes. 
		The soft modes along $\Gamma \rightarrow$R are artifacts of k-point 
		interpolation. Their contributions to the DoS is negligible and they do 
		not affect the electron-phonon coupling results, which account only for 
		vibrational modes commensurate with the Born-von~Karman simulation cell.
	The vibrational dispersion and DoS of Ic are very similar to those of Ih.}
	\label{fig:Phonon_BandstructureAndDoS}
\end{figure}
The vibrational density of states (DoS) of Ih (shown in Fig.\ 
\ref{fig:Phonon_BandstructureAndDoS}) and Ic splits into five regimes:
pseudo-translation modes (I), librational modes (II), molecular bending
modes (III) and molecular (anti-) symmetric O-H stretching modes (IV
and V).  In Ref.\ \cite{monserrat} we showed that crystal modes
(pseudo-translation and librational modes) and molecular modes (bending
and O-H stretching modes) each contribute around half of the large
band-gap correction in Ih $P6_{3}cm$ of around $\Delta E_g$(0)
$\approx -1.5 \eV$ obtained for a 192-molecule simulation cell. 

\begin{figure}
	\centering
	\includegraphics[width=0.4\textwidth]{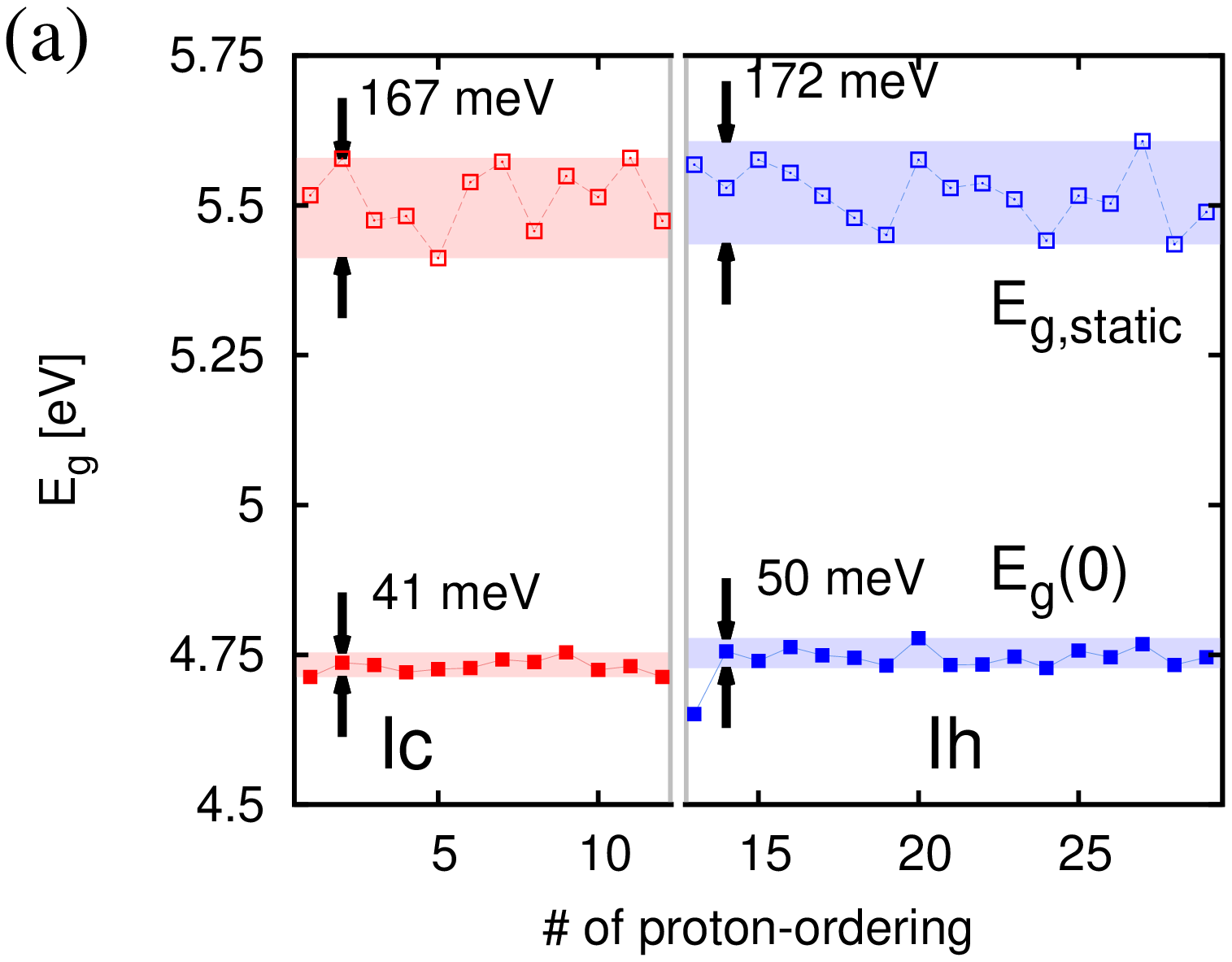}
	\includegraphics[width=0.4\textwidth]{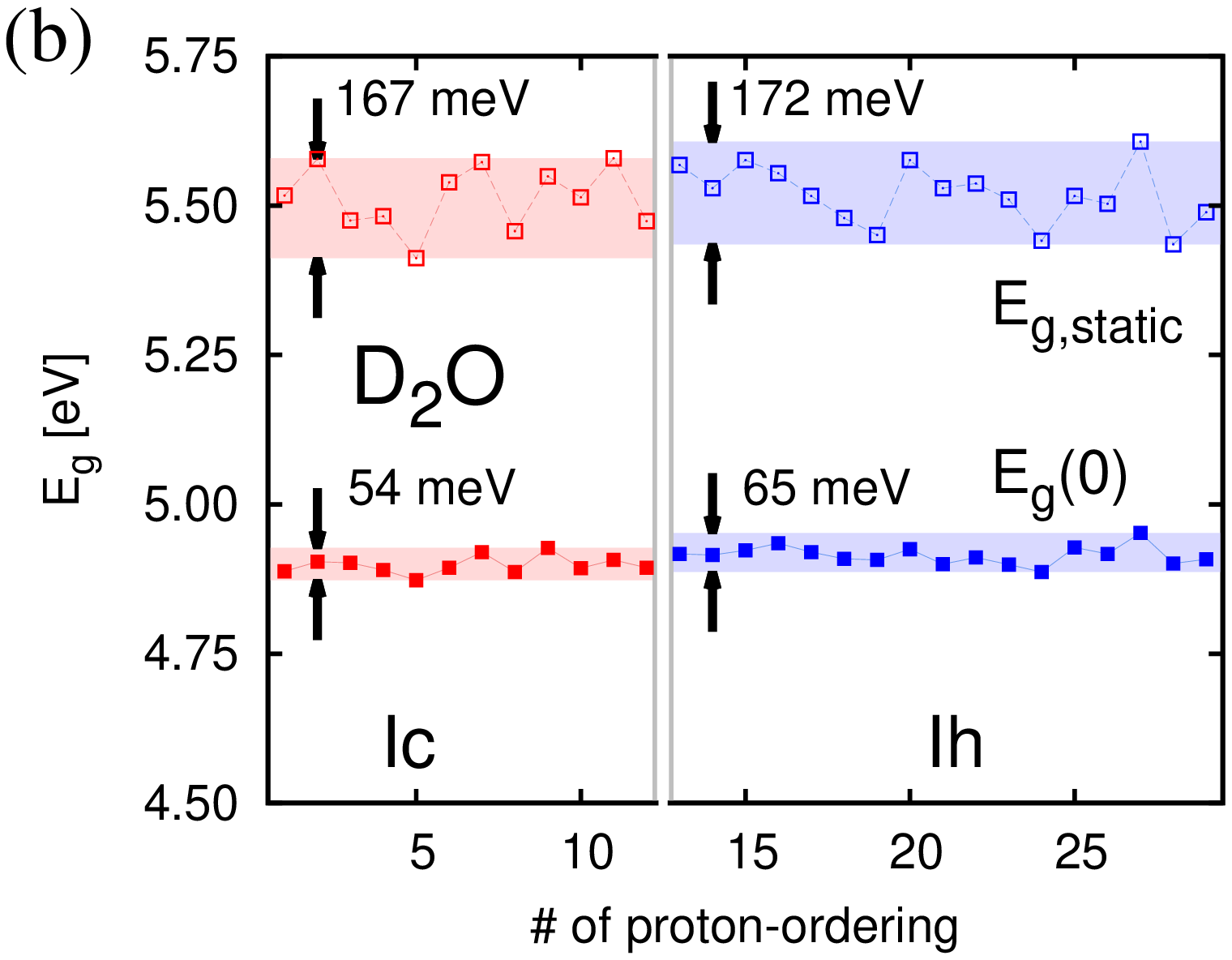}
	\caption{Static-lattice band gaps, $E_{g,\textrm{static}}$, and 
		$E_{g}$(0) of 8-molecule (a) H$_2$O and (b) D$_2$O proton-orderings. 
		The change in the band gap due to vibrations ($\Delta E_{g}$) of Ih 
		and Ic are very similar. $\Delta E_{g}$(0) for D$_2$O is significantly 
		smaller than for H$_2$O. 
		The $\Delta E_{g}$ for Ih $P6_{3}cm$ has been extrapolated down to a 
		cell size of 8 molecules by fitting a 2nd order polynomial to the data 
		points for 12, 96 and 192 molecules to compare it with the $\Delta E_{g}$ 
	of the 8-molecule Ih and Ic proton-orderings.}
	\label{fig:Polytypes_BandGaps}
\end{figure}

Figs.~\ref{fig:Polytypes_BandGaps} (a) and (b) show $\Delta E_{g}$(0) 
for protiated and deuterated 8-molecule proton-ordered Ih and Ic structures.
The band-gap corrections are very similar in Ih and Ic.
While the vibrational corrections, $\Delta E_{g}$, shown in Fig.\
\ref{fig:Polytypes_BandGaps} (a) obtained for the 8-molecule unit
cells are only converged to simulation cell size errors of around 
$0.75\eV$, the relative $\Delta E_{g}$ of the proton-orderings are 
converged to less than $75 \meV$ (see Section \ref{Methods} and, in particular,
Fig.\ \ref{fig:SC_Conv}), which is sufficient to distinguish between 
different proton-orderings.

Figs.\ \ref{fig:Polytypes_BandGaps} (a) and (b) show that nuclear vibrations 
largely average out differences in the band gaps obtained with different 
proton-orderings, reducing variations across proton-orderings in $E_{g}$(0) 
by almost 3/4, i.e., from around $170 \meV$ to $\leq 50\meV$.  
This occurs because the protons are effectively smeared out by vibrational motion.
The root-mean-square (RMS) vibrational displacements of the protons of around 
$0.23 \ang$ correspond to about 1/3 of the difference in the bond length of an 
O-H covalent bond and an O$\cdots$H hydrogen bridge bond. While we did not remark 
on this in our earlier publication \cite{engel}, this smearing effect also occurs 
in the free energies of proton-orderings, albeit in a far less striking way than 
for the band gaps. 
See Supplementary Section III \cite{supplementary} and Fig.\ S5 in particular for 
further details.
\begin{figure}
	\centering
	\includegraphics[width=0.4\textwidth]{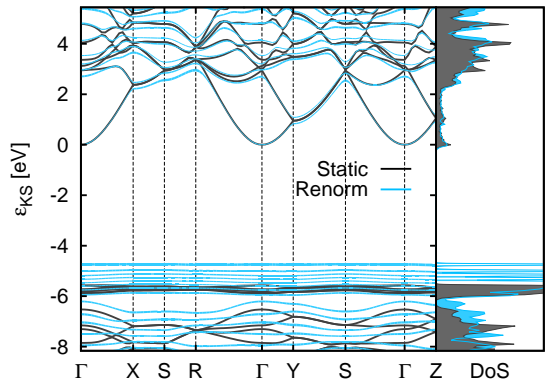}
	\caption{Renormalisation of electronic bandstructure and DoS for 8-molecule 
		Ih $Cmc2_1$. The conduction band minimum (CBM) has been set to zero energy. 
		The broadening of the peak in the DoS at the top of the valence band indicates 
		strong coupling of the phonons to the valence band maximum (VBM). Vice versa, 
		the minimal broadening of the peaks at the bottom of the conduction band 
		indicate weaker coupling of the phonons to the CBM. This picture remains 
		if the energy of the lowest Kohn-Sham (KS) state is chosen as the reference 
	instead of the CBM.}
	\label{fig:Bandstructure_Renorm}
\end{figure}
The full static lattice and vibrationally renormalised bandstructures 
for 8-molecule Ih $Cmc2_1$ are shown in Fig.\ \ref{fig:Bandstructure_Renorm}.

\begin{figure}
	\centering
	\includegraphics[width=0.4\textwidth]{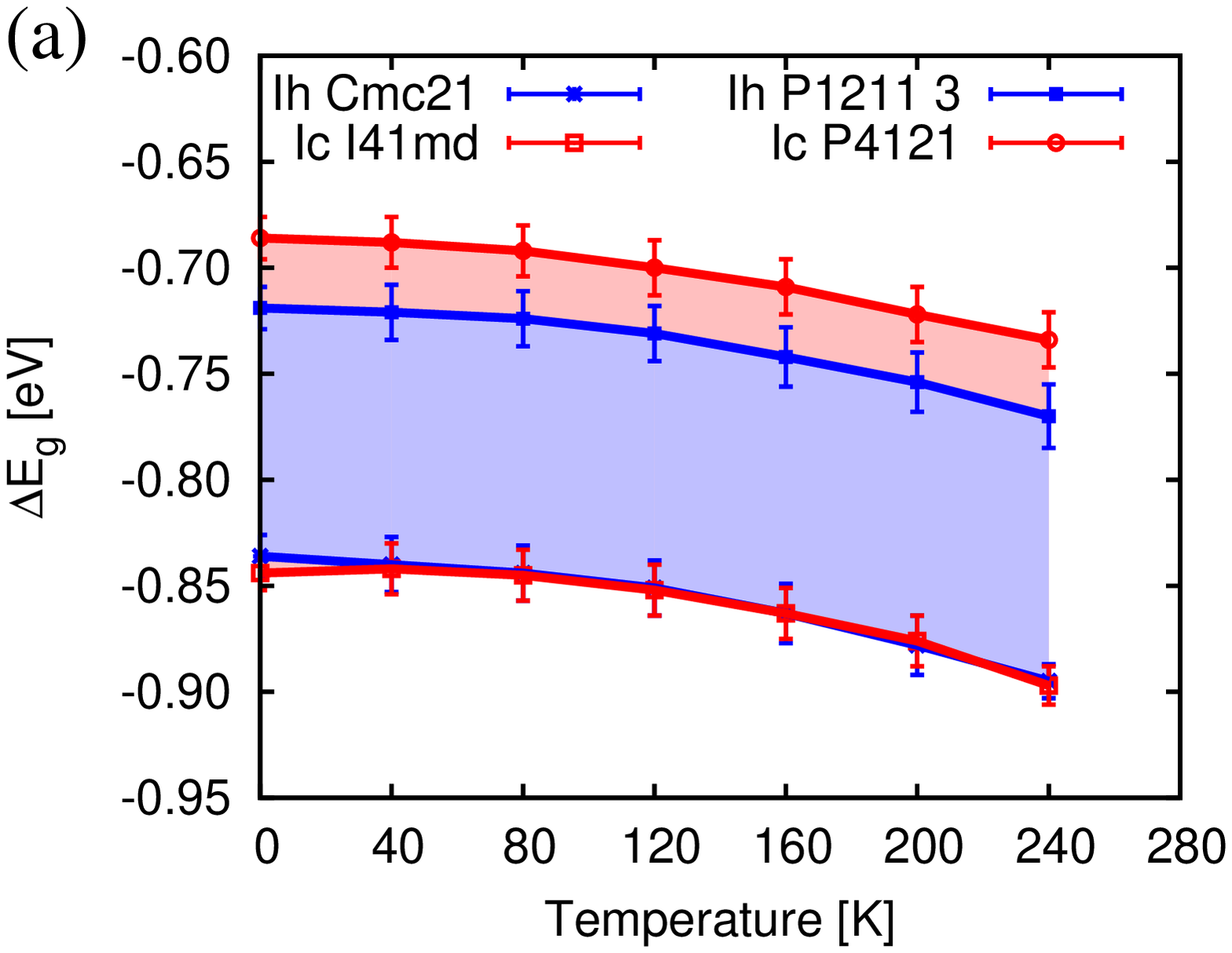}
	\includegraphics[width=0.4\textwidth]{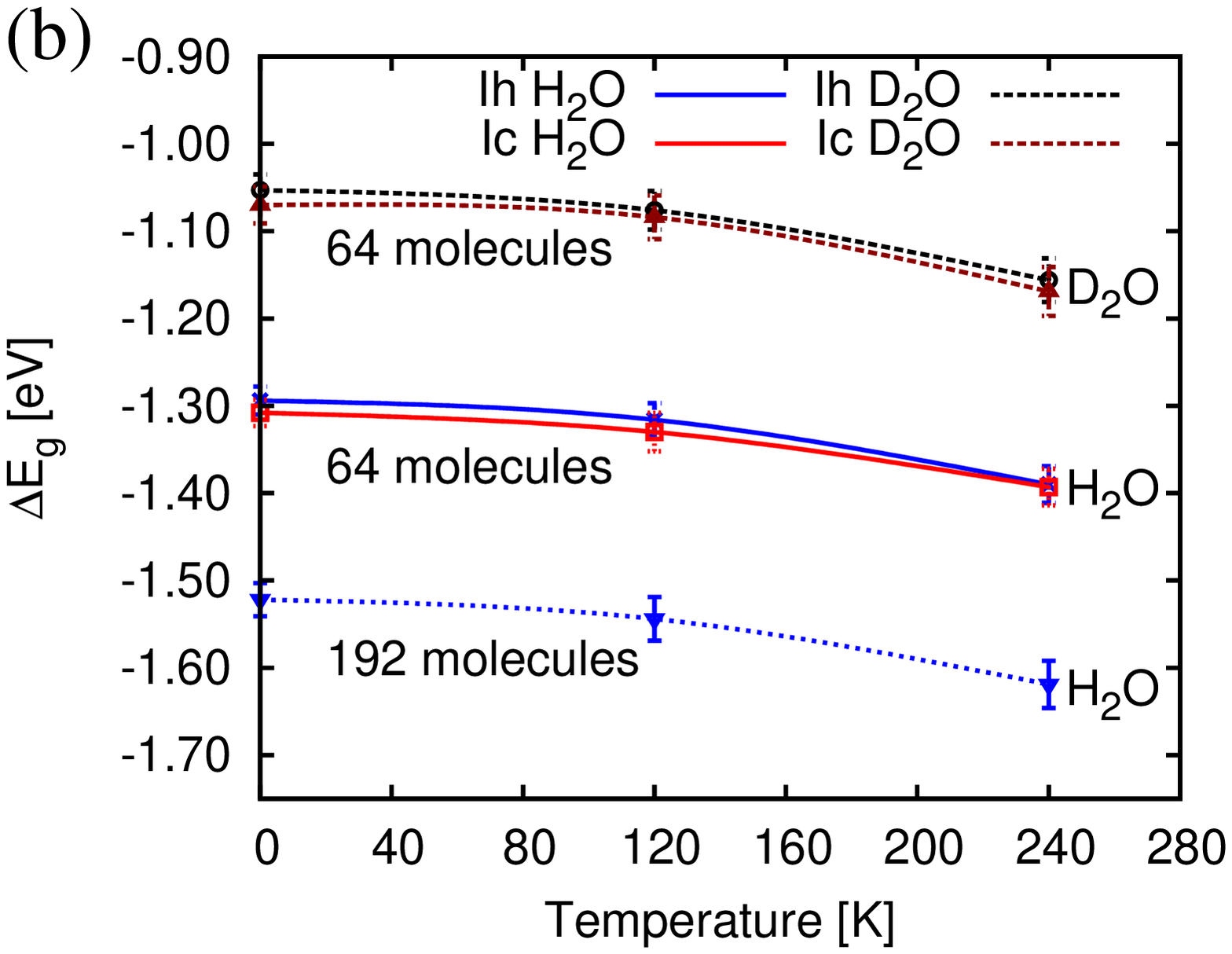}
	\caption{(a) Temperature dependence of $E_{g}(T)$ for the protiated 8-molecule 
		Ih and Ic proton-orderings with the largest and smallest values of 
		$\Delta E_{g}$(0), respectively. (b) Temperature dependence of 
		$E_{g}(T)$ of 64-molecule H$_2$O and D$_2$O proton-orderings. 
		$\Delta E_{g}(T)$ in Ih and Ic are very similar and the temperature 
		dependence is just as strong in deuterated as in protiated ice. The 
		most converged result for the 192-molecule Ih $P6_3cm$ structure is 
	shown as a blue dotted line.}
	\label{fig:Polytypes_TempDep}
\end{figure}
We find that the temperature dependence of $E_{g}(T)$ is almost
independent of the proton-ordering. 
Fig.\ \ref{fig:Polytypes_TempDep} (a) shows the temperature 
dependence of $E_{g}$ for 8-molecule simulation cells of the Ih and 
Ic proton-orderings with the smallest and largest $\Delta E_{g}$(0), 
respectively.
In ice, even up to the melting
temperature of $273$ K, only the lowest frequency crystal vibrational
modes (up to $24 \meV$) are thermally activated. These are long-wavelength 
modes that do not probe the differences between proton-orderings. As can 
be seen in Fig.\ \ref{fig:BGapCorrVsFreq}, these low-energy crystal modes 
contribute less than one tenth of the ZP band-gap correction and accordingly 
the thermal correction up to melting at 273 K of around $95\meV$ is less than 
one tenth of the quantum ZP band-gap correction, $\Delta E_g$(0). Notably,
Ref.\ \cite{shibaguchi} reports the experimental observation of a shift in 
the (excitonic gap) of $85 \pm 5 \meV$ upon heating from 103 K to 254 K, 
which is in reasonable agreement with our calculated value of around $80\meV$ 
obtained for the 192-molecule Ih $P6_3cm$ simulation cell (see Fig.\ 
\ref{fig:Polytypes_TempDep}).

\section{Deuterated Ice}
\label{HeavyIce}

Deuteration of ice leads to a significant softening of all but the low
frequency crystal vibrational modes. The resulting vibrational DoS is
shown in panel (a) of Supplementary Fig.\ S8.  The larger nuclear mass 
of deuterons in comparison to protons leads to smaller vibrational
displacements. This isotope effect leads to a smaller ZP $\Delta
E_{g}$(0), as shown in Fig.~\ref{fig:Polytypes_BandGaps} (b) 
for 64-molecule simulation cells.
At the same time, the smaller deuteron vibrational displacements result
in a smaller smearing out of the deuteron positions and thus a clearer
distinction in $E_{g}$(0) between different proton-orderings.  The
variations across proton-orderings increase from around 41--50 $\meV$ 
in protiated ice to 54--65 $\meV$ in deuterated ice.  The thermal 
contribution to $\Delta E_{g}(T)$, $\Delta_{0 \mathrm{\rightarrow} T} 
E_{g} \equiv \Delta E_{g}(T) - \Delta
E_{g}$(0) in deuterated ice, however, is of comparable size to that 
in protiated ice (see Fig.~\ref{fig:Polytypes_TempDep} (b)). This 
follows from the very similar forms of the low frequency DoS of protiated 
and deuterated ice shown in panel (c) of Supplementary Fig.\ S8. The low 
frequency crystal modes up to frequencies of $24 \meV$, which become 
thermally activated at temperatures below 273 K, are collective modes 
for which the effect of the heavier deuterons is masked by the even 
heavier oxygen atoms.  
At high temperatures one might expect the higher ZP $\Delta E_{g}$(0) 
in protiated ice to be reflected in a larger temperature dependence, 
but the crossover point at which all modes are thermally activated is 
far above melting.

\section{The Role of Vibrational Anharmonicity}
\label{Anharmonicity}

In Ref.~\cite{engel} we showed that the effect of (typically quartic) 
anharmonicity in the Born-Oppenheimer energy surface is to localise the 
nuclear wavefunctions by 1.5\% and 2.5\% with respect to the harmonic 
wavefunctions of Ih and Ic, respectively. 
This can be understood as a reduction of the anharmonic vibrational 
displacements, $q^{\textrm{anh}}_{n{\bf k}}$, with respect to the harmonic 
vibrational displacements, $q^{\textrm{har}}_{n{\bf k}}$, by a factor 
$\alpha = 0.985$ and $\alpha = 0.975$ for Ih and Ic, respectively:
$q^{\textrm{anh}}_{n{\bf k}} = \alpha q^{\textrm{har}}_{n{\bf k}}$. 
Since the dependence of the band gap on $q_{n,{\bf k}}$ is typically 
well approximated by $\Delta {\hat E}_{g}(q_{n{\bf k}}) \approx 
c^{(1)}_{n{\bf k}} \vert q_{n{\bf k}} \vert + c^{(2)}_{n{\bf k}} \vert q_{n{\bf k}} \vert^2$
(see Fig.\ \ref{fig:FailureHarmApprox}) we expect the anharmonic 
band-gap correction in Ih,
$\Delta {\hat E}^{\textrm{anh}}_{g}(q_{n{\bf k}}) \approx 
\alpha \left( c^{(1)}_{n{\bf k}} \vert q^{\textrm{har}}_{n{\bf k}} \vert 
+ \alpha c^{(2)}_{n{\bf k}} \vert q^{\textrm{har}}_{n{\bf k}} \vert^2 \right)$,
to be between $\alpha$ of the harmonic value, 
$\Delta E^{\textrm{har}}_{g}$, if $c^{(1)}_{n{\bf k}} \gg c^{(2)}_{n{\bf k}}$ 
to $\alpha^2$ of the harmonic value if $c^{(2)}_{n{\bf k}} \gg c^{(1)}_{n{\bf k}}$.
Correspondingly, we expect anharmonic corrections to the band gap 
of +1.5\% to +2.25\% of the $\Delta E^{\textrm{har}}_g$.
Similarly we expect corrections in Ic of +2.5\% to +6.25\% of the harmonic 
$\Delta E_g$.
The harmonic band-gap correction is around $-1.5 \eV$ for Ih and Ic, which 
translates into expected zero-temperature band-gap corrections due to 
anharmonicity of around 20--35 $\meV$ and 40--90 $\meV$ for Ih and Ic, 
respectively.
Using the reweighting procedure described in Section\ \ref{ElPhCoupling}, 
we calculate anharmonic corrections to $E_g$(0) of +$33 \pm 18 \meV$ and 
+$76 \pm 21 \meV$ for Ih and Ic, respectively, in good agreement with 
the estimates given above. We use the 96-molecule Ih $P6_3cm$ and the 
64-molecule Ic $P4_3$ simulation cells.

We note that the band gap corrections due to vibrational anharmonicity 
in Ih and Ic, and different proton-orderings, only differ by around $20 \meV$. 
Hence the variations in $E_{g}$ across proton-orderings remain small at 
around $50 \meV$, independent of whether anharmonic nuclear motion is included.

\section{Atomistic Insights}
\label{Mechanistics}

\begin{figure}
	\centering
	\includegraphics[width=0.4\textwidth]{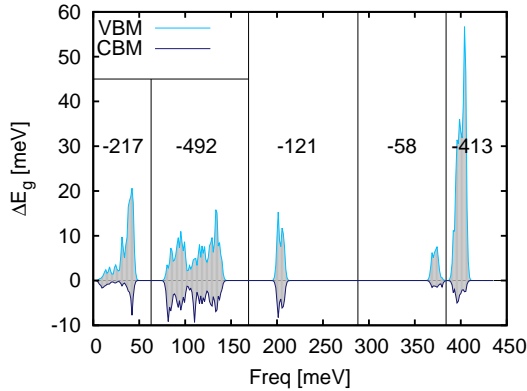}
	\caption{Contributions of vibrational modes to $\Delta E_{g}$(0) for 
		64-molecule Ih $Cmc2_1$. 
		Unlike the band gap and, consequently, the vibrational corrections to it, 
		the individual vibrational corrections to the KS energies of the VBM and 
		CBM cannot be measured, but can be understood in terms of the VBM and CBM 
	electronic densities shown in Fig.\ \ref{fig:ElectronDensities}.}
	\label{fig:BGapCorrVsFreq}
\end{figure}
As shown in Fig.\ \ref{fig:BGapCorrVsFreq}, the pseudo-translations 
(0--50 $\meV$), molecular bending modes (195--214 $\meV$) and anti-symmetric 
O-H bond stretching modes (363--384 $\meV$) together contribute less than $1/3$ 
of $\Delta E_{g}$(0)  (the individual contributions are around 16\%, 10\% and 
4\%, respectively). Librational modes (76--150 $\meV$) and symmetric O-H bond 
stretching modes (387--411 $\meV$) give large contributions of around $38 \%$ 
and $32 \%$, respectively.  

\begin{figure}
	\includegraphics[width=0.225\textwidth]{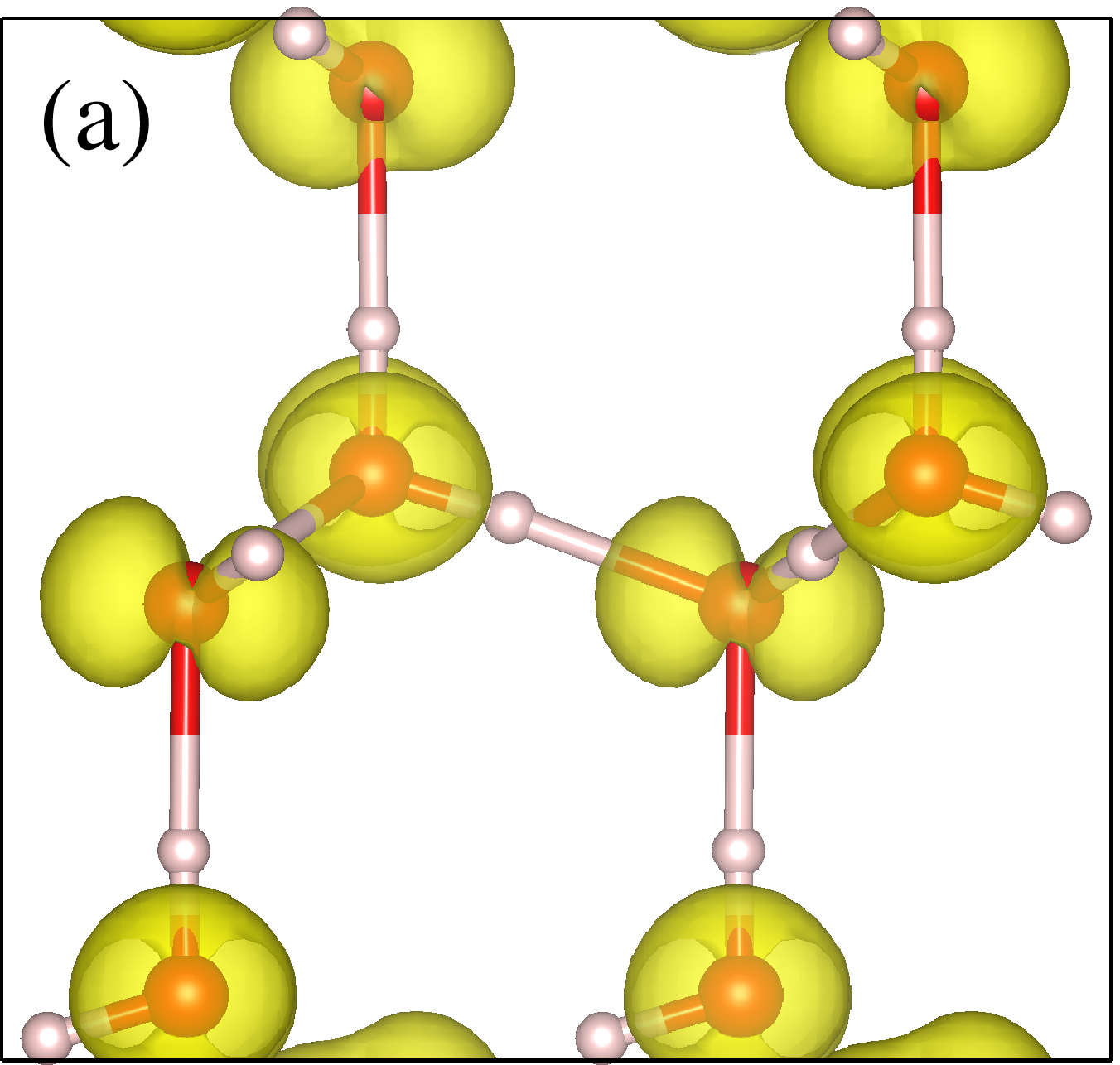}
	\includegraphics[width=0.225\textwidth]{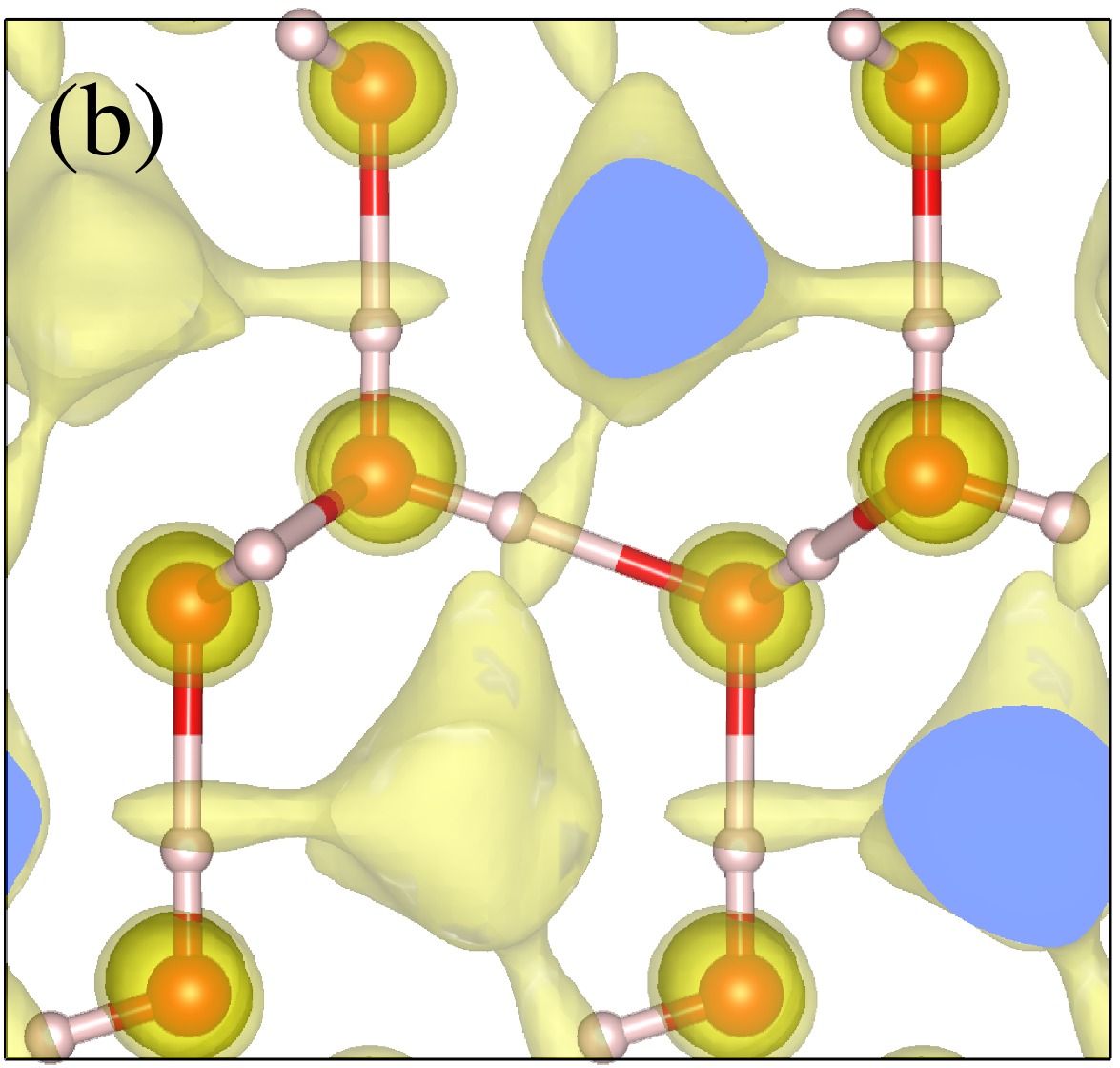}
	\caption{(a) VBM and (b) CBM electronic densities of Ih $Cmc2_1$.
		The electronic charge density of the VBM is localised in oxygen
		lone-pair-like orbitals which form the hydrogen bridge network.
		Thus the VBM couples strongly to the molecular and librational
		modes. The CBM is delocalised (pale yellow isosurface), but the
		bulk of the electron density is localised in orbitals of
		anti-bonding $\sigma^\star$ character, centered on the oxygen
	nuclei.}
	\label{fig:ElectronDensities}
\end{figure}

Overall, the vibrational motion of the nuclei predominantly couples to the 
valence band maximum (VBM), as shown in Fig.\ \ref{fig:Bandstructure_Renorm}, 
in which the peaks corresponding to the valence band states are strongly 
renormalised and broadened, while the peaks corresponding to the conduction 
band states remain relatively sharp and distinct.
The electron density corresponding to the VBM sits in distorted oxygen lone pair 
orbitals (see Fig.~\ref{fig:ElectronDensities} (a)).
Consequently, the electronic density corresponding to the VBM is significantly 
distorted by displacements of the protons along their O-H covalent and O$\cdots$H 
bridge bonds. Such displacements arise, in particular, from the symmetric O-H 
bond stretching modes, which therefore strongly affect the Kohn-Sham (KS) energy 
of the VBM.

Librational modes also couple to the VBM, albeit less strongly. Unlike the 
O--H bond stretching modes, librational modes additionally couple significantly 
to the conduction band minimum (CBM).
The electron density corresponding to the CBM is predominantly localised in 
orbitals of anti-bonding $\sigma^\star$ character centered on the oxygen nuclei 
(see Fig.\ \ref{fig:ElectronDensities} (b)). These are almost identical for 
different proton-orderings, due to their practically identical oxygen sublattices 
(up to a difference in stacking of layers between Ih and Ic). Consequently the 
electron density corresponding to the CBM is very similar for all proton-orderings. 
Unlike the O-H bond stretching modes, the librational modes distort the oxygen
sublattice with respect to the static configuration, momentarily (though not
on average) breaking the tetrahedral coordination of the oxygen atoms and
affecting the electron density forming the CBM.
Nonetheless, the coupling of the vibrational motion of the nuclei to the CBM 
is weak due to the atomic-like nature of the electron density of the CBM.

Besides leading to the largest contribution to $\Delta E_{g}$(0), O--H stretching 
modes also play the predominant role in smearing out the distinction between 
proton-orderings, thereby making the VBM energies of different proton-orderings 
more similar. This effect of reducing differences in $\Delta E_{g}$(0) between 
proton-orderings is clearly shown in Fig.\ \ref{fig:Polytypes_BandGaps}.
While the strong renormalisation of the KS energy of the VBM leads to strong 
vibrational renormalisation of $E_{g}$, the free energy (which corresponds to 
the sum over all vibrationally renormalised occupied KS bands) is much less 
affected. Consequently the narrowing of the distribution of lattice free energies 
of different proton-orderings is far less pronounced than for the band gaps (see 
Supplementary Section III).

\section{Conclusions}
\label{Conclusions}

Accurate static lattice electronic band gaps can be 
calculated using theoretical methods such as DFT with semi-local or 
hybrid functionals, the $GW$ method, or highly accurate quantum 
chemical methods.
However, it is important to account for vibrational effects when
benchmarking calculated electronic band gaps against experiment. We have
found vibrational corrections to the static electronic band gaps of 
water ice of $-1.5$ to $-1.7 \eV$, which are significant on the scale 
of the differences between results obtained with these methods and 
compared to the size of the experimental gap.

Proton-disorder, on the other hand, does not play an important role 
in determining the electronic bandstructure of Ih and Ic.  The use of
different proton-orderings in an atomistic simulation affects the 
energetics of Ih and Ic on the scale of the difference in free energy 
between Ih and Ic.  However, when vibrations are accounted for, the 
renormalised zero and finite temperature electronic bandstructures of 
different proton-ordered Ih and Ic structures are very similar. 
Hence, atomistic simulations of band gaps and optical absorption 
spectra depend only weakly on the proton-ordering when vibrations 
are taken into account.
The similarity of the vibrationally renormalised electronic bandstructures 
of different Ih and Ic proton-orderings also implies that measurements 
of the band gaps or optical absorption spectra of ice samples are not 
capable of distinguishing between proton-orderings or between Ih, Isd 
and Ic.
Comparing experimental band gaps of protiated and deuterated ice 
would provide a straightforward means of accessing quantum zero-point 
effects.
Such measurements might allow the resolution of the predicted difference 
in the band gaps of protiated and deuterated ice of around $0.2 \eV$.

Non-quadratic behaviour plays a crucial role in water ice. For example, 
accurate vibrationally renormalised electronic band gaps can only be 
calculated by carefully accounting for the strongly non-quadratic 
dependence of the band gap on the vibrational displacements. In fact, 
if accuracies of tens of $\meV$ for electronic band gaps are to be 
achieved, vibrational anharmonicity must be accounted for as well. 
Anharmonic nuclear vibrations are crucial in understanding the 
relative stability of Ih and Ic \cite{engel} and many other phenomena.
It would be of great interest to study the role of 
anharmonicity at ice surfaces and in the presence of impurities or other 
defects.

\vspace{0.25cm} {\bf Acknowledgements.}  We acknowledge financial support 
from the Engineering and Physical Sciences Research Council of the UK 
[EP/J017639/1]. 
B.\ M.\ also acknowledges Robinson College, Cambridge, and the Cambridge
Philosophical Society for a Henslow Research Fellowship.
The calculations were performed using the Cambridge High
Performance Computing Service facility and the Archer facility of the
UK's national high-performance computing service (for which access was
obtained via the UKCP consortium [EP/K013564/1]).


%

	\end{document}